\documentclass[%
 reprint,
 showpacs,preprintnumbers,
 amsmath,amssymb,
 aps,
 prb,
]{revtex4-1}
\usepackage{dcolumn}
\usepackage[dvipdfmx]{graphicx}
\usepackage{subfigure}
\usepackage[dvipdfmx]{color}
\usepackage{mathrsfs}

\makeatletter
\def\btt#1{\texttt{\@backslashchar#1}}
\DeclareRobustCommand\bblash{\btt{\@backslashchar}} \makeatother

\begin{document}

\title{Tunneling Conductance in a Two-dimensional Dirac Semimetal Protected by Non-symmorphic Symmetry }

\author{Tetsuro Habe}
\affiliation{Department of Physics, Osaka University, Toyonaka, Osaka 560-0043, Japan}

\date{\today}

\begin{abstract}
We theoretically study a tunneling effect in a two-dimensional Dirac semimetal with two Dirac points protected by non-symmorphic symmetries.
The tunnel barrier can be arranged by a magnetic exchange potential which opens a gap at the Dirac points which can be induced by a magnetic proximity effect of a ferromagnetic insulator. 
We found that the tunnel decay length increases with a decrease in the strength of the spin-orbit coupling, and moreover the dependence is attributed to the correlation of sublattice and spin degree of freedoms which lead to symmetry-protected Dirac points.
The tunnel probability is quite different in two Dirac points, and thus the tunnel effect can be applied to the highly-selective valley filter.
\end{abstract}


\maketitle
Dirac semimetal has the gapless energy spectrum of electrons with a point node at which the conduction and the valence bands are touched and such a node, so-called Dirac point, emerges at a symmetrical point in the Brillouin zone\cite{Murakami2007,Wan2011,Burkov2011,Neupane2013,Borisenko2013}.
The electric excitation energy has a linear dependence on the wave number from the Dirac point at which the even number of electric states are degenerated.
One can control the low energy spectrum by a symmetry-breaking external field and change the transport phenomena.

In recent years, several kinds of candidates and theoretical predictions of the realistic Dirac semimetal in three-dimensions have been proposed\cite{Wang2012,Young2012,Borisenko2014,Liu2014}.
The wealth of candidates in three-dimensions is associated with the simple necessary condition for realization of the Dirac semimetal which requires the band inversion and crystal symmetry $C_3$.
In two-dimensions, on the other hand, there are few candidate compounds with Dirac points.
Graphene had been regarded as a two-dimensional(2D) Dirac semimetal and studied as a test ground for researching the unique features of massless Dirac fermions\cite{Neto2009}.
However, the linear dispersion in graphene is an approximation even in a clean system because the spin-orbit interaction opens an energy gap and makes it to be a topological insulator\cite{Kane2005,Kane2005-2}.

Recently, it was proposed that non-symmorphic symmetries are necessary for Dirac points to be stable in a 2D system\cite{Young2015}, and these symmetries also play an important role in one- and three-dimensional semimetals\cite{Zhao2016,Schoop2016}.
In practice, it is shown that the Dirac points in such a 2D Dirac semimetal are preserved even in the presence of the spin-orbit interaction\cite{Liu2016}.
%
However, the spin-orbit interaction in graphene is quite small $\sim2K$, and thus it seems that there is no difference in ordinary experimental observations between graphene and the symmetry-protected 2D Dirac semiemtal.
Thus it is important to suggest the different physical property of the electrons in the symmetry-protected Dirac semimetal from the nearly massless excitation in graphene.

In this paper, we discuss the tunneling electric transport in the symmetry-protected 2D Dirac semimetal with two Dirac points, i.e. the valley degree of freedom.
The tunnel barrier is fabricated by a magnetic exchange potential, e.g. which is induced by the magnetic proximity effect in a junction with a ferromagnetic insulator.
We show that the perpendicular component of the exchange field opens a gap in the 2D Dirac semimetal similar to the surface states of three-dimensional topological insulators\cite{Yokoyama2010,Habe2012} or the sublattice-symmetry-breaking potential in graphene\cite{Neto2009}.
However, the exchange potential, unlike that in the three-dimensional topological insulator, is not a simple mass term for massless Dirac fermions in the 2D Dirac semimetal because of the spin-sublattice correlation associated with non-symmorphic symmetry.
We find that the tunnel decay length, unlike ordinary 2D Dirac fermions, becomes longer with a decrease in the spin-orbit coupling because of the sublattice-spin correlation in the symmetry-protected 2D Dirac semimetal, and propose that this tunneling system plays a role of the highly-selective valley filter.

We consider a model Hamiltonian proposed by Young and Kane for describing electric states in the symmetry-protected 2D Dirac semimetal with a square lattice including two atoms in a unit cell\cite{Young2015}, and the simplest form is given by,
\begin{align*}
H_0=&2t\tau_x\cos\frac{k_x}{2}\cos\frac{k_y}{2}+t_2(\cos k_x+\cos k_y)\\
&+t_{so}\tau_z(\sigma_y\sin k_x-\sigma_xk_y),
\end{align*}
where $\tau$ and $\sigma$ are Pauli matrices in the sublattice and spin spaces, respectively.
The electric states have three Dirac points $M=(\pi,\pi)$, $X_1=(\pi,0)$, and $X_2=(0,\pi)$ in the first Brillouin zone. 
The energy of node at $X_1$ and $X_2$ is equal to each other but it is different from that at $M$ because of the chiral symmetry breaking term $t_2$, and thus there is no dip of density of states in this model.
However, the $C_2$ screw symmetry breaking term, which is introduced to simulate iridium oxide superlattice proposed by Chen and Kee\cite{Chen2014},
\begin{align*}
V_1=\Delta_1\sin\frac{k_x}{2}\sin\frac{k_y}{2}\tau_x,
\end{align*}
opens a gap only at the $M$ point.
The Hamiltonian $H=H_0+V_1$ represents a rigorous 2D Dirac semimetal.

The electric transport is associated with the electric states around the Fermi energy, and thus we can discuss the transport phenomenon for the Fermi energy near the Dirac node by using the effective Hamiltonian based on $kp$ theory around the $X_j$ as 
\begin{align*}
H_{X_1}(p)=&-t_{so}\tau_z(\sigma_yp_x+\sigma_xp_y)+\tau_x(-tp_x+\Delta_1p_y)\\
H_{X_2}(p)=&t_{so}\tau_z(\sigma_yp_x+\sigma_xp_y)+\tau_x(-tp_y+\Delta_1p_x),
\end{align*}
with a relative wave vector $\boldsymbol{p}=(p_x,p_y)$ from the $X_j$.
In what follows, we discuss the tunneling effect at only the $X_1$, however the result is applicable to the case at the $X_2$ with $t_{so}\rightarrow-t_{so}$, $-t\rightarrow\Delta_1$, and $\Delta_1\rightarrow-t$.

\begin{figure}[htbp]
\begin{center}
 \includegraphics[width=75mm]{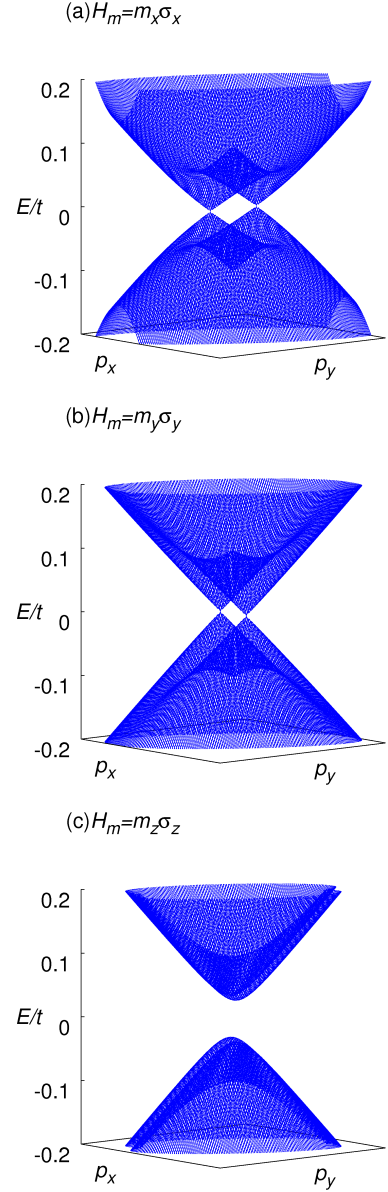}
\caption{The energy dispersion of the 2D Dirac semimetal around $X_1$ with a magnetic moment $H_m=m_x\sigma_x$(a), $m_y\sigma_y$(b), and $m_z\sigma_z$ (c). 
 }\label{Band_with_magnet}
\end{center}
\end{figure}

First, we consider the effect of a magnetic exchange potential on the electric states in the symmetry-protected 2D Dirac semimetal and we show that the potential enables us to open the energy gap at the Dirac node.
For instance, such an exchange potential $m_\mu\sigma_\mu$ can be induced by a magnetic proximity effect of the junction with a ferromagnetic insulator, and the coupling constants $m_\mu$ can be controlled by changing the magnetization in the ferromagnetic insulator.
To analyze the effect of the exchange potential, we rewrite the potential in the basis of the eigenvectors for $H_{X_1}$,
\begin{align*}
U_{\theta,\phi}^\dagger H_{X_1}U_{\theta,\phi}=\sqrt{t_{so}^2p^2+(-tp_x+\Delta_1p_y)^2}\tau_z\sigma_z,
\end{align*}
where the unitary operator $U_{\theta,\phi}$ consists of two rotation matrices $R_{\sigma,\theta}$ in the spin subspace and $R_{\tau,\phi}$ in the sublattice subspace,
\begin{align*}
U_{\theta,\phi}=R_{\sigma,\theta}\frac{1}{\sqrt{2}}(1+i\tau_x\sigma_z)R_{\tau,\phi},
\end{align*}
with $\boldsymbol{p}=(p\cos\theta,p\sin\theta)$, $\sin\phi=(-tp_x+\Delta_1p_y)/\varepsilon_0$, and
\begin{align*}
R_{\sigma,\theta}=\frac{1}{\sqrt{2}}(\sigma_z+\sigma_y\cos\theta+\sigma_x\sin\theta).
\end{align*}
In this basis, the spin operators are transformed into
\begin{align*}
{U}_{\theta,\phi}^\dagger \sigma_zU_{\theta,\phi}=&(-\sigma_x\cos\theta+\sigma_y\sin\theta)\\
&\times(\tau_z\sin\phi-(\tau_x\cos\phi+\tau_y\sin\phi)\cos\phi)\\
U_{\theta,\phi}^\dagger \sigma_xU_{\theta,\phi}=&\sigma_z\sin\theta
-\cos\theta(\sigma_y\cos\theta+\sigma_x\sin\theta)\\
&\times(\tau_z\sin\phi-(\tau_x\cos\phi+\tau_y\sin\phi)\cos\phi)\\
U_{\theta,\phi}^\dagger \sigma_yU_{\theta,\phi}=&\sigma_z\cos\theta
+\sin\theta(\sigma_y\cos\theta+\sigma_x\sin\theta)\\
&\times(\tau_z\sin\phi-(\tau_x\cos\phi+\tau_y\sin\phi)\cos\phi).
\end{align*}
One can find a particular angle $\theta_0$ for any exchange potential coupling to an in-plane spin to be the identity in the sublattice space, and thus the in-plane component of the exchange field $m_\mu$ preserves the gapless energy dispersion where two Weyl nodes can be found on the line $\boldsymbol{p}$ with this angle $\theta_0$.
The potential coupling to the out-of-plane spin $m_z\sigma_z$, on the other hand, opens a gap in the energy spectrum because it provides non-zero component proportional to $\tau_\mu\sigma_\nu$ even with any angles $\theta$ and $\phi$.
We show the energy dispersion in the presence of the magnetic exchange potential coupling to out-of-plane spin and in-plane spin in Fig.\ref{Band_with_magnet}.
The effect of the in-plane exchange field is similar to the separation of the Dirac node into two Weyl nodes in the time-reversal breaking Weyl semimetal with a exchange potential\cite{Habe2014}.

\begin{figure}[htbp]
\begin{center}
 \includegraphics[width=75mm]{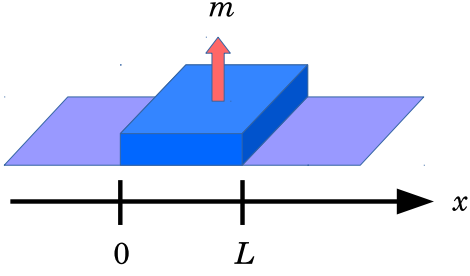}
\caption{The schematic picture of the tunneling junction arranged by a ferromagnetic insulator. 
 }\label{Tunneling_junction}
\end{center}
\end{figure}

Next, we consider the tunnel junction arranged by the exchange potential $m\sigma_z$ in the symmetry-protected 2D Dirac semimetal where the tunneling barrier can be fabricated by attaching a ferromagnetic insulator locally as shown in Fig.\ref{Tunneling_junction}.
The junction system can be described by 
\begin{align}
H=&H_{\xi}(0)\theta(-x)+H_{\xi}(m)\theta(x)\theta(L-x)\notag\\&+H_{\xi}(0)\theta(x-L),\label{effective_Hamiltonian}
\end{align}
where the Hamiltonian can be represented in the basis of the eigenvectors for the glide mirror operator $\tau_x\sigma_z$,
\begin{align}
H_\xi(m)=&\left(\xi m+tp_x-\Delta_1p_y\right)s_z-t_{so}(\xi s_yp_x+s_xp_y),\label{Local_Hamiltonian}
\end{align}
with a Pauli matrix $s$ and the eigenvalue of the glide mirror operator $\xi=\tau_x\sigma_z$ because the glide mirror symmetry is preserved even in the presence of $m\sigma_z$.
Here, the Pauli matrix $s$ is the pseudo spin in the basis of eigenstate for the glide mirror operator and it represents the staggered alignment of spin at two sublattices in each spin axis.
The eigenstate with the incident electron with the energy $\varepsilon$ for the Hamiltonian (\ref{effective_Hamiltonian}) can be written by a wave function consisting of three functions smoothly connected at the boundaries of the three regions,
\begin{align*}
\Psi=\begin{cases}
\psi_{\xi,p_x^+}^+e^{ip_x^+x}+R\psi_{\xi,p_x^-}^+e^{ip_x^-x}&(x<0)\\
C_1\psi_{\xi,q_x^+}^+e^{iq_x^+x}+C_2\psi_{\xi,q_x^-}^+e^{iq_x^-(x-L)}&(0<x<L)\\
T\psi_{\xi,p_x^-}^+e^{ip_x^+(x-L)}&(L<x)
\end{cases},
\end{align*}
where $\psi_{\xi,p_x}^\pm$ is the eigenfunction of Eq.(\ref{Local_Hamiltonian}).
Here, $T$ and $R$ are the transmission and reflection coefficients.
If the two boundaries of the second region $0<x<L$ are assumed to be parallel to each other, the wave number parallel to the interface is preserved in the scattering, and the wave number perpendicular to the boundary is a function of $m$ as $p_x^\pm=k^\pm(m)$ and $q_x^\pm=k^\pm(0)$ with
\begin{align}
k^\pm(m)=&-\frac{t(\xi m-\Delta_1p_y)}{t^2+t_{so}^2}\notag\\
&\pm\frac{\sqrt{(t^2+t_{so}^2)(\varepsilon^2-t_{so}^2p_y^2)-t_{so}^2(\xi m-\Delta_1p_y)^2}}{t^2+t_{so}^2},\label{wave_number}
\end{align}
for the eigenstate with the energy $\varepsilon$.
The exchange potential provides a tunnel barrier to the electrons with the energy $|\varepsilon|<t_{so}\sqrt{p_y^2+(\xi m-\Delta_1p_y)^2/(t^2+t_{so}^2)}$. 

\begin{figure}[htbp]
\begin{center}
 \includegraphics[width=75mm]{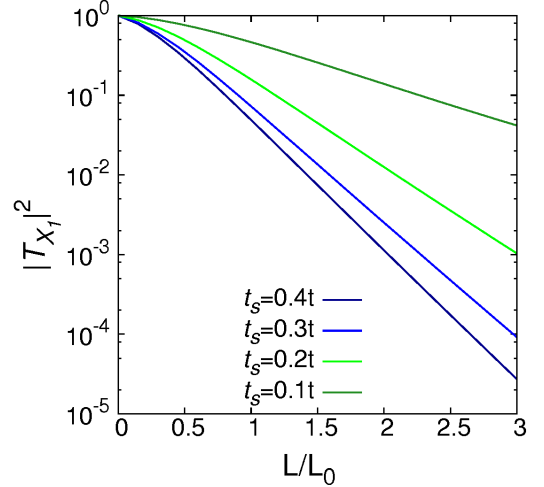}
\caption{The tunneling probabilities at $X_1$ as a function of $L$ with $\varepsilon_F=0.02t$, $\Delta_1=0.5t$, $m=0.2t$, and $L_0=t/\varepsilon_F$. 
 }\label{Tunneling1}
\end{center}
\end{figure}

In the tunnel junction, the analytic formulation of the transmission coefficient $T(p_y)$ can be obtained by smoothly connecting the wave functions at the boundaries $x=0$ and $x=L$, and it can be represented by
\begin{align}
T(p_y)^{-1}=&-e^{-iq_x^+L}\left(1-\frac{\gamma(m,q_x^-,0,p_x^-)}{\gamma(0,p_x^+,0,p_x^-)}\frac{\gamma(0,p_x^+,m,q_x^+)}{\gamma(m,q_x^-,m,q_x^+)}\right.\notag\\&\left.\times(1-e^{-i(q_x^--q_x^+)L})\right),\label{Transmission_coefficient}
\end{align}
with
\begin{align*}
\gamma(m_1,k_1,m_2,k_2)=&(\varepsilon+m_1+tk_1-\Delta_1p_y)(p_y+i\xi k_2)\\
&-(\varepsilon+m_2+tk_2-\Delta_1p_y)(p_y+i\xi k_1),
\end{align*}
for each channel of $p_y$.
We show the tunneling probability as a function of the length of the barrier region $L$ in Fig.\ref{Tunneling1}.
One can find that the mean value of the tunneling probability $|T_{X_1}|^2$ shows the typical property of an ordinary tunnel junction where the tunneling probability decreases with an increase in $L$.

However, the damping factor $\kappa=\mathrm{Im}[q_x^+]$ shows the characteristic feature of the symmetry-protected 2D Dirac semimetal unlike the ordinary 2D Dirac fermion.
The damping factor can be written by a function of the ratio $r_{so}$ of the spin-orbit coupling constant $t_{so}$ to the hopping matrix $t$,
\begin{align*}
\kappa=\sqrt{\frac{r_{so}^2(\xi m/t-\Delta_1/tp_y)^2}{(1+r_{so}^2)^2}-\frac{\varepsilon^2/t^2-r_{so}^2p_y^2}{(1+r_{so}^2)}},
\end{align*}
and the tunnel decay length $\kappa^{-1}$ drastically increases with a decrease in the spin-orbit coupling. 
This is because the insulating gap induced by the magnetic exchange potential reduces with a decrease in the spin-orbit interaction.
This dependence of the gap on the spin-orbit coupling constant is quite different from the case of the ordinary 2D Dirac fermion where the gap is fixed by the mass term, i.e. the exchange potential, and does not change with the strength of the spin-orbit coupling.
The extension of the tunnel decay length with the spin-orbit coupling can be observed by controlling the strength $t_{so}$ which can be realized, e.g. by the effect of substrates.

\begin{figure}[htbp]
\begin{center}
 \includegraphics[width=75mm]{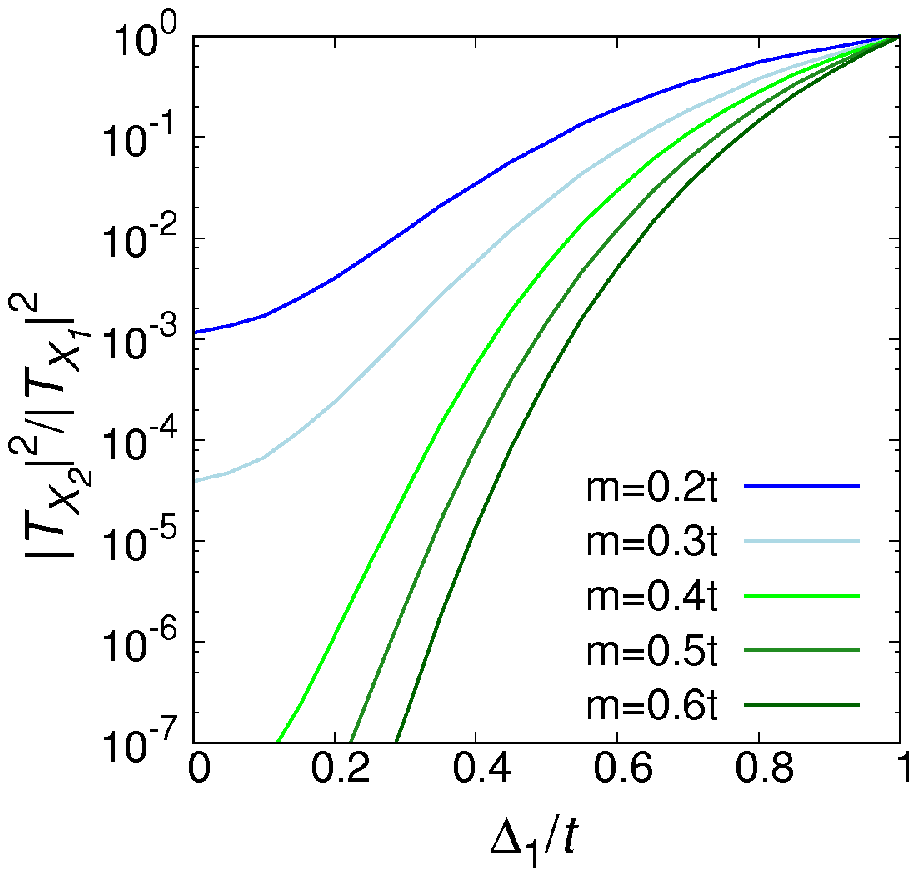}
\caption{The ratio of tunneling probabilities at $X_1$ and $X_2$ as a function of $\Delta_1/t$ with $\varepsilon_F=0.02t$ and $t_{so}=0.4t$. 
 }\label{Tunneling2}
\end{center}
\end{figure}

\begin{figure}[htbp]
\begin{center}
 \includegraphics[width=75mm]{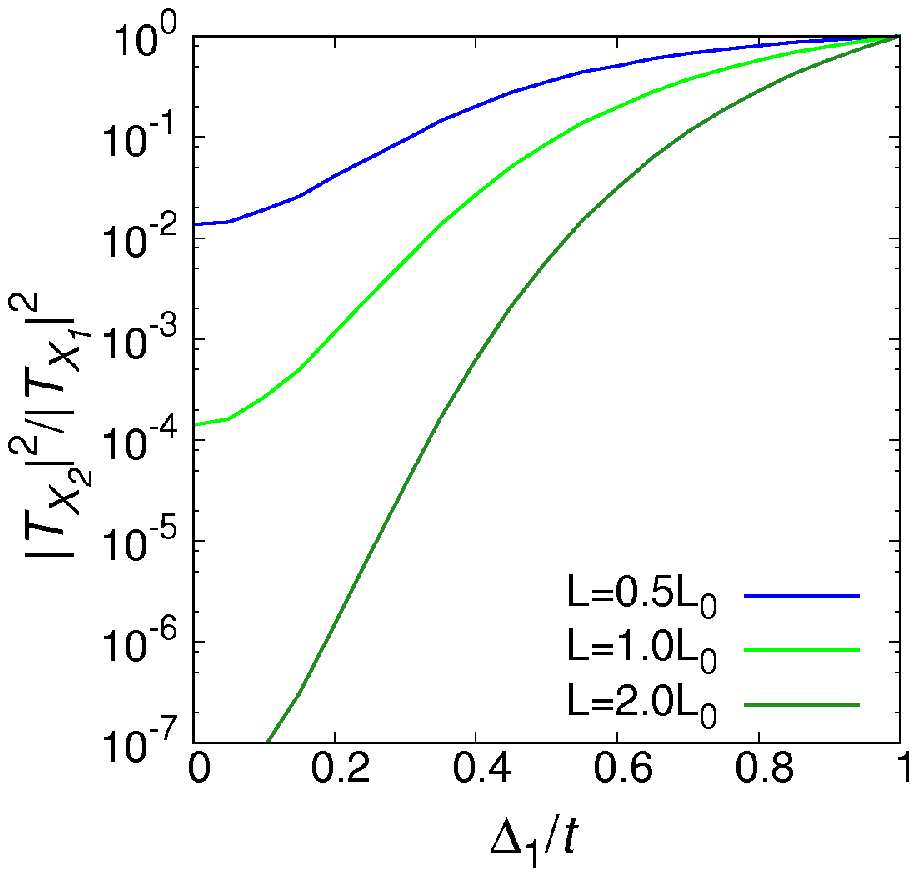}
\caption{The ratio of tunneling probabilities at $X_1$ and $X_2$ as a function of $\Delta_1/t$ with $\varepsilon_F=0.02t$ and $t_{so}=0.4t$. 
 }\label{Tunneling3}
\end{center}
\end{figure}

Finally, we discuss the difference between two Dirac points $X_1$ and $X_2$ in the tunneling effect.
We show the ratio of tunneling probabilities at two Dirac points as a function of the relative strength of distortion of the lattice $\Delta_1/t$, which describes the non-equivalence in the two points, in Fig.\ref{Tunneling2} and \ref{Tunneling3}. 
When the distortion is small $\Delta_1/t\ll1$, the tunneling at the $X_2$ point are negligibly small compared with that at the $X_1$ point.
This is because the sublattice degree of freedom works as a pseudo spin and the reflection is suppressed at the $X_1$ point by a mismatch of the pseudo spin between incident and reflected waves.
In Eq.(\ref{Local_Hamiltonian}), the direction of the pseudo spin is determined by two factors; the hopping matrix and the spin-orbit interaction. 
The hopping matrix between the sublattces with the hopping parameters $t$ and $\Delta_1$ couples to $s_z$ in the subspace, and the spin-orbit interaction is represented by the in-plane component of the pseudo spin $s_x$ and $s_y$.
Therefore, the pseudo spin is nearly aligned in the $z$ direction because the spin-orbit interaction is generally much smaller than the hopping matrix.
With a small distortion $\Delta_1/t\ll1$, the pseudo spin of the incident wave with $0<p_x$ and the reflected wave with $p_x<0$ is nearly anti-parallel at the $X_1$.
The incident and reflected waves at the $X_2$, on the other hand, have the nearly parallel pseudo spin because the sign of $p_y$, which is preserved in the tunneling process, is relevant to its direction.
In practice, the tunneling probability is same in two valleys under the condition of $\Delta_1/t=1$ where the contribution of $p_x$ to the alignment of the pseudo spin is unchanged between the two points.

The difference of tunneling probability in two valleys can be enhanced with an increase in the exchange potential $m$ and the length of the insulating region as shown in Fig.\ref{Tunneling2} and \ref{Tunneling3}, respectively.
The dominant valley for the tunneling can be selected by the direction of the tunneling junction.
The asymmetric tunneling effect in two valleys produces the valley polarized current and gives a way to control the valley degree of freedom without valley Hall effect. 

In conclusion, we studied the tunneling effect in the non-symmorphic symmetry-protected 2D Dirac semimetal with a tunneling barrier arranged by the magnetic exchange potential, and found that the tunneling decay length shows a quite different feature from ordinary 2D Diac fermions as a function of the strength of the spin-orbit interaction.
The characteristic property is attributed to the $C_2$ screw symmetry-breaking interaction which preserves the other non-symmorphic symmetry about a glide mirror operation and induces a rigorous 2D Dirac semimetal.
We also found that the tunneling junction works as a highly selective valley filter.

\bibliography{2D_Dirac}

\end{document}